\documentclass[a4paper,11pt]{article}
\usepackage{jheppub} 
\usepackage{lineno}

\usepackage[english]{babel}
\usepackage{indentfirst}
\usepackage{amsmath}
\usepackage{amsfonts}
\usepackage{physics}
\usepackage{tensor}
\usepackage{bbm}
\usepackage{graphicx}
\usepackage{tabularx}

\usepackage[colorinlistoftodos]{todonotes}

\title{On correlators in $T\Bar{T}$-deformed conformal field theories}

\author[a,b]{D. Menskoy}
\affiliation[a]{Moscow Institute of Physics and Technology, 141700, Dolgoprudny, Russia}
\affiliation[b]{Kharkevich Institute for Information Transition Problems, Russian Academy of Sciences, 127994, Moscow, Russia}

\emailAdd{menskoy.dd@phystech.edu}

\abstract{This paper is our contribution to the study of $T\bar{T}$-deformations. We consider the effect of $T\bar{T}$-deformation of conformal field theories in perturbation theory. We use dimensional regularization scheme to perturbatively calculate the effect of the $T\bar{T}$-deformation on the correlation functions of the CFT's primary fields. We provide the renormalization of the fields for the case of two-point correlation function.}

\makeatletter
\gdef\@fpheader{}
\makeatother

\begin{document}
\maketitle
\flushbottom

\section{Introduction}
\label{sec:intro}

Recently, considerable attention has been paid to the study of $T\bar{T}$-deformations of quantum field theories (\cite{Zamolodchikov_2004expectation}, \cite{Smirnov_2017}, \cite{Cavagli_2016}). Increased interest to this topic was caused by the fact that for a given deformation there are analytical formulas that characterize physical quantities, despite the irrelevant nature of this perturbation. Among these quantities are the bulk energy density, partition function on a torus (\cite{Cardy_2018}, \cite{Datta_2018}) and the scattering matrix (\cite{Dubovsky_2017}, \cite{Smirnov_2017}, \cite{Cavagli_2016}). The $T\bar{T}$-deformation is determined by the equation for the effective action


\begin{equation}
    \frac{d}{d\alpha}\mathcal{A}_\alpha = \int (T\bar{T})_\alpha(x)\: d^2x,
\end{equation}
where $T\bar{T}$ is a specially selected composite operator. It was shown in \cite{Zamolodchikov_2004expectation}, that on cylinder the energy and momentum spectrum satisfy the Burgers equation
\begin{equation}\label{burgers}
    \pdv{E_n}{\alpha} + E_n \pdv{E_n}{R} + \frac{P_n^2}{R} = 0.
\end{equation}

Studying the correlation functions in $T\Bar{T}$-deformed quantum field theories is problem of a special interest. It was studied with the help of UV-cutoff regularization method in \cite{Cardy_2019}, \cite{He_2020}, \cite{He_2022note}, \cite{He_2023systematic} and with the help of dimensional regularization in \cite{Giribet_2020} and \cite{Dey_2021}.

One can try to solve the $T\bar{T}$-flow equation perturbatively. For the case, when $\mathcal{A}_{\alpha = 0 } $ is the action of conformal field theory one has
\begin{equation}
    \mathcal{A}_\alpha = \mathcal{A}_0 + \alpha \int (T\bar{T})_0(x)\: d^2x\: + \alpha^3\int d^2x \: (L_{-2}\bar{L}_{-2}T\bar{T})_0(x)+ \mathcal{O}(\alpha^4),
\end{equation}
where we omitted the terms that are total derivatives. It is expected that the term of order $\alpha^3$ (i.e. $\alpha^3\int d^2x \: (L_{-2}\bar{L}_{-2}T\bar{T})_0(x)$) will remove all occuring third order divergencies that appeared due to perturbation $\alpha \int (T\bar{T})_0(x)\: d^2x$. In this work we make some steps in the direction of checking this.

This paper is organized as follows. In Section \ref{sec:criticality} we remind the connection of $T\bar{T}$-deformation with subleading singularities in statistical systems. In Section \ref{section_corr} we study corrections to correlation function of several primary fields in $T\Bar{T}$-deformed conformal field theories. In Subsection \ref{sec:first_order} we calculate the first-order correction and renormalize the fields. In Subsection \ref{sec:second_order} we make some calculations that substitutes the second-order correction, calculate the divergences in two-point correlation function and renormalize the fields.

\section{$T\bar{T}$-deformation and critical phenomena}\label{sec:criticality}

Let us consider a two-dimensional statistical system in the vicinity of the critical temperature $T_c$. $T\bar{T}$-deformation describes the nature of subleading singularities in statistical physics. The free energy of such a system behaves as
\begin{align}
    F = F_0 + a (T - T_c)^{2\nu} + a' (T - T_c)^{\Xi} + ... \: ,
\end{align}
where $\nu$ and $\Xi$ are some exponents, connected with relevant and least irrelevant operators in effective action, $F_0$ is a function regular for $T \to T_c$, and "..."\:denotes less singular contributions. In this case, the behavior of the correlation radius $R_c = M^{-1}$ is as follows:
\begin{align}
    M = b (T - T_c)^{\nu} + b'(T - T_c)^{\eta} + ...
\end{align}

The nature of singularities can be understood in terms of the Wilson renormalization group. The behavior directly at the critical point is determined by the fixed point of the renormalization group (conformal field theory). In the neighborhood of a fixed point, the system is described by the formal action
\begin{align}
    \mathcal{A} = \mathcal{A}_{CFT} + \mu \int d^2x \:\Phi_{\Delta}(x) + \sum_{i} \alpha_i \int d^2x \: O_i(x),
\end{align} 
where $\mathcal{A}_{CFT}$ is the action of conformal field theory, $2 \Delta < 2$ and, therefore, the perturbation $\mu \int d^2 x \:\Phi_{\Delta}(x)$ is relevant, the remaining terms are irrelevant contributions, with the dimensions of the corresponding operators $d_i > 2$. For simplicity, the case of one relevant perturbation is considered. It is also assumed that there are no marginal operators.

It is important that the coupling constants are regular functions of the temperature. Also, since $\Phi_{\Delta}$ is a relevant perturbation, then $\mu$ should vanish at $T = T_c$:
\begin{align}
    \mu (T) &= \mu' (T - T_c) + \mu'' (T - T_c)^2 + ...
    \\
    \alpha_i &= \alpha_i^{(0)} + \alpha_i' (T - T_c) + ...
\end{align}
That means that there is a “canonical curve”\:in the space of quasilocal actions\footnote{By quasilocal action we mean such actions 
\begin{equation}
    \mathcal{A} = \int d^2x \:\mathcal{L}(\Phi(x), \partial_\mu\Phi(x), \partial_\mu\partial_\nu\Phi(x), ...),
\end{equation}
that the Lagrangian density is a convergent functional for all fields, whose Fourier expansion consists of modes with $|k| < \Lambda$ only with $\Lambda$ being a UV-cutoff. } that describes the dependence of the microscopic theory on the temperature. For $T = T_c$, the renormalization group flow drives the thoery from a point on this curve strictly to the fixed point. For $T - T_c \ll T_c$ the renormalization group flow first flows similarly towards a fixed point, then stays for long 'RG time' in the vicinity of a fixed point and then flows away towards some other fixed point. All critical singularities are formed in the vicinity of a fixed point, and irrelevant contributions die out there. Renormalization group equations at this stage
\begin{align}
    \dv{\mu}{l} = (2 - 2\Delta) \mu, \: \dv{\alpha_i}{l} = (2 - d_i) \alpha_i.
\end{align}
And since the correlation length behaves as $R_c (l) \sim e^{-l}$ with $l$ -- infinitesimal change in rescaling factor, then we find the leading singularities
\begin{align}
    R_c \sim \mu ^{-\nu} \sim \abs{T - T_c}^{-\nu}, \: \nu = \frac{1}{2 - 2\Delta}.
\end{align}
 Amplitudes $a$ and $b$ are determined from the scattering matrix of quantum field theory \cite{Klassen_1991}
 \begin{align}
    \mathcal{A}_{CFT} + \mu \int d^2 x \:\Phi_{\Delta}(x).
 \end{align}
 This phenomenon is known as the universality of the leading critical singularity.
 
 The following subleading singularities reflect the influence of the least irrelevant operator, that is, one such that
 \begin{align}
    d_1 < d_i, \: i \geq 2.
 \end{align}
 In many systems (for example in Ising model), such an operator, respecting all symmetries of the system, is $T\Bar{T}$. From dimensional considerations we obtain
 \begin{align}
    \Xi = d_1 \nu, \: \eta = (d_1 - 1) \nu.
 \end{align}
In the case of $T\Bar{T}$ one has
 \begin{align}
    d_{T\Bar{T}} = 4, \: \Xi = 4 \nu, \: \eta = 3 \nu.
 \end{align}
 For amplitudes $a'$ and $b'$ the phenomenon of universality is violated. However, the statement about partial universality remains true, namely
 \begin{equation}
    \frac{a}{b} = \frac{a'}{b'}.
 \end{equation}
 We can derive that as follows. From the equation (\ref{burgers}) we can obtain the following dependences for the correlation radius and free energy \cite{Smirnov_2017}
 \begin{align}
     F_{\alpha} = \frac{F(0)}{1 + \alpha F(0)}, \: F(0) = F_{\alpha = 0} = a \mu^{2\nu},
     \\
     M_{\alpha} = \frac{M(0)}{1 + \alpha F(0)}, \: M(0) = M_{\alpha = 0} = b \mu^{\nu}.
 \end{align}
 Then to first order in $\alpha$ we have \cite{Zamolodchikov_lectures}
 \begin{align}
    F_{\alpha} = F(0) - \alpha F(0)^2 + \mathcal{O}(\alpha^2),
    \\
    F_{\alpha} = M(0) - \alpha M(0)F(0) + \mathcal{O}(\alpha^2).
 \end{align}
 Thus
 \begin{align}
     a' = -\alpha F(0)^2, \: b' = - \alpha F(0)M(0).
 \end{align}
 Now for the ratio $a'/b'$ we obtain the required relation
 \begin{align}
     \frac{a'}{b'} = \frac{-\alpha F(0)^2}{-\alpha F(0) M(0)} = \frac{F(0)}{M(0)} = \frac{a}{b}.
 \end{align}

\section{Correlation functions in $T\bar{T}$-deformed theory} \label{section_corr}

We will study at the quantum level the solution to the $T\bar{T}$-flow equation
\begin{equation}
     \pdv{\mathcal{A}_\alpha}{\alpha} = \int d^2x (T\bar{T})_\alpha (x)
\end{equation}
under the initial condition $\mathcal{A}_0 = \mathcal{A}_{CFT}$ .
This equation has a solution in the form of a series
\begin{equation}
     \mathcal{A}_\alpha = \mathcal{A}_0 + \alpha \int d^2x (T\bar{T}) (x) + \mathcal{O}(\alpha^3),
\end{equation}
where there are no terms of order $\alpha^2$ due to dimension considerations\footnote{In fact, operator $\partial \Bar{\partial} (T\Bar{T}$) satisfy the dimensional condition in order $\alpha^2$, but gives no contribution as a total derivative}. Thus, it is necessary to consider in perturbation theory the correlation functions of the theory
\begin{equation}
     \mathcal{A}_\alpha = \mathcal{A}_0 + \alpha \int d^2x (T\bar{T}) (x).
\end{equation}
\subsection{First order correction} \label{sec:first_order}

We will calculate the correlation functions of primary fields. In first order we need to calculate the integral
\begin{equation}\label{eq:first-ordre_int}
     \int d^2 z \expval{T\Bar{T}(z) \Phi_1(x_1) \Phi_2(x_2)... \Phi_N(x_N)}_{0},
\end{equation}
where $z = z^1 + i z^2$, $d^2 z = dz^1 dz^2$, the subscript '0' denotes that the correlation functions are calculated in the unperturbed theory, i.e. in conformal field theory. Correlation functions of primary fields with the stress-energy tensor in conformal field theory can be expressed through the ones of the primary fields. In particular, the integrand in (\ref{eq:first-ordre_int}) has the form
\begin{equation}
     \sum_{j,k = 1}^{N}\left(\frac{\Delta_j}{(z-x_j)^2} + \frac{1}{z-x_j}\pdv{}{x_j} \right) \left(\frac{\Bar{\Delta}_k}{(\bar{z} - \bar{x}_k)^2} + \frac{1}{\bar{z} -\bar{ x}_k}\pdv{\bar{x}_k}\right)\expval{\Phi_1(x_1)...\Phi_N(x_N)}_0,
\end{equation}
where $\Delta_j$ and $\bar{\Delta}_j$ are the conformal and anticonformal dimensions of the primary field $\Phi_j$.

As we will see further only the integrals of the form
\begin{equation}
     I_m^n (x, y) = \int d^2 z \frac{1}{(z-x)^m(\bar{z} - \bar{y})^n}
\end{equation}
with $m,n = 1,2$ arise in the first order. These integrals diverge and need to be regularized. We will use dimensional regularization, that is, we consider the analytic continuation of integrals as functions of dimension $d = 2 + \epsilon$. In this approach, integrals of the form $I_m^n(x,x)$ are postulated to be zero. 

One can notice that as
\begin{multline}
     I_m^n(y,x) = \int dz^1 dz^2 \frac{1}{(z^1+i z^2-y)^m(z^1-i z^2-\bar{x} )^n}=\\ = -\int dz^1 dz^2 \frac{1}{(z^1-i z^2-y)^m(z^1+i z^2-\bar{x} )^n} = -I_n^m(\bar{x}, \bar{y}),
\end{multline}
then it is enough to compute only $I_1^1(x,y)$, $I_2^1(x,y)$, $I_2^2(x,y)$. Here we just quote the results of these and all other emerging integrals in \ref{table:ints}.

\begin{table}\label{table:ints}
\centering
\begin{tabular}{|c|c|c|}
 \hline
 $m$ & $n$ & $I_m^n(2+\epsilon)$ \\
 \hline
 1 & 1 &  $-\frac{2\pi}{\epsilon} - \pi\log{\bigl(|x-y|^2 m_R^2\bigr)} + \mathcal{O}(\epsilon)$ \\
 \hline
 2 & 1  & $-\frac{\pi}{x-y} - \frac{\pi\log(m_R |x-y|) }{x-y}\epsilon + \mathcal{O}(\epsilon^2)$  \\
\hline
 2 & 2  & $0 + \frac{\pi}{2|x-y|^2} \epsilon + \mathcal{O}(\epsilon^2)$  \\
 \hline
 3 & 1  & $-\frac{\pi}{2(x-y)^2}  + \mathcal{O}(\epsilon)$  \\
 \hline
  3 & 2  & $0  + \mathcal{O}(\epsilon)$  \\
 \hline
  4 & 1  & $-\frac{\pi}{3(x-y)^2}  + \mathcal{O}(\epsilon)$  \\
 \hline
  4 & 2  & $0  + \mathcal{O}(\epsilon)$  \\
 \hline
\end{tabular}
\caption{List of integrals' expansions in the limit $d=2+\epsilon\to2$. Here $m_R^2 = \pi e^{\gamma_E-1}$.}
\end{table}

We can now collect all the terms together and calculate the original integral
\begin{multline}
     \int d^2 z \expval{T\Bar{T}(z) \Phi(x_1) \Phi(x_2)... \Phi(x_N)}_{0} =
     \\
     =\int d^2z \sum_{j,k = 1}^{N}\left(\frac{\Delta_j}{(z-x_j)^2} + \frac{1}{z-x_j}\pdv {}{x_j} \right) \left(\frac{\Bar{\Delta}_k}{(\bar{z} - \bar{x}_k)^2} + \frac{1}{\bar{ z} -\bar{x}_k}\pdv{\bar{x}_k} \right)\expval{\Phi_1(x_1)...\Phi_N(x_N)}_0
     \\
     =\sum_{j=1}^{N}\sum_{k\neq j}^{N}\biggl(\frac{\pi}{\bar{x}_j-
     \bar{x}_k}\bar{\Delta}_k \partial_{j} - \frac{\pi}{x_j-x_k}\Delta_j\bar{\partial}_k + \biggl(-\frac{2\pi}{\epsilon}- 2\pi \log{(|x_j-x_k|m)}\biggr)\partial_j \bar{\partial}_k \biggr) \expval{\Phi_1(x_1)...\Phi_N(x_N)}_0
\end{multline}
Let us transform the last term using the Ward identity for correlation functions in conformal field theory
\begin{align}
     \sum_{i=1}^{N} \partial_{i}\expval{ \Phi_1(x_1) \Phi_2(x_2)... \Phi_N(x_N)}_{0} &= 0,
     \\
     \sum_{i=1}^{N} \bar{\partial}_{i}\expval{ \Phi_1(x_1) \Phi_2(x_2)... \Phi_N(x_N)}_{0} &= 0.
\end{align}
Using these identities we can rewrite the last term as
\begin{equation}
    \begin{split}
        \sum_{j=1}^{N}\sum_{k\neq j}^{N} \biggl(&-\frac{2\pi}{\epsilon}- 2\pi \log{(|x_j-x_k |m)}\biggr)\partial_j \bar{\partial}_k \expval{\Phi_1(x_1)... \Phi_N(x_N)}_0 =
         \\
         &= \sum_{j=1}^{N}\biggl(-2\pi\left(\log m+\frac{1}{\epsilon}\right)\sum_{k = 1}^N \partial_j \bar {\partial}_k + 2\pi \left(\log m+\frac{1}{\epsilon}\right)\partial_j \bar{\partial}_j
         \\
         &-2 \pi \sum_{k\neq j}^{N} \log{|x_j-x_k|}\partial_j \bar{\partial}_k\biggr) \expval{\Phi_1(x_1)... \Phi_N(x_N)}_0
         \\
         &= \biggl(2\pi\left(\log m+\frac{1}{\epsilon}\right)\sum_{j=1}^{N}\partial_j \bar{\partial}_j - 2\pi \sum_{j=1}^{N}\sum_{k\neq j}^{N} \log{|x_j-x_k|}\partial_j \bar{\partial}_k\biggr) \expval{\Phi_1(x_1)... \Phi_N(x_N)}_0
    \end{split}
\end{equation}
As a result we have a first-order correction regularizes answer
\begin{equation}
    \begin{split}
    \biggl[\sum_{j=1}^{N}\sum_{k\neq j}^{N}\biggl(\frac{\pi}{\bar{x}_j-
     \bar{x}_k}\bar{\Delta}_k \partial_{j} - \frac{\pi}{x_j-x_k}\Delta_j\bar{\partial}_k - 2\pi \log{(|x_j-x_k|)}\partial_j \bar{\partial}_k \biggr)+
     \\
     +2\pi\left(\log m+\frac{1}{\epsilon}\right)\sum_{j=1}^{N}\partial_j \bar{\partial}_j\biggr]\expval{\Phi_1(x_1)...\Phi_N(x_N)}_0
    \end{split}
\end{equation}

\subsubsection{Renormalization}

Now we can perform the renormalization. Namely: let’s combine fields in such linear combinations
\begin{equation}
     \Phi^{R}_i (x_i) = \Phi_i(x_i) + \alpha A(\epsilon) \pdv{x_i}\pdv{\bar{x}_i}\Phi_i(x_i)
\end{equation}
and calculate for them in the first order the correlation functions of the perturbed theory
\begin{equation}
     \expval{\Phi_1^R(x_1)...\Phi_N^R(x_N)}_{def} = \expval{\Phi_1^R(x_1)...\Phi_N^R(x_N)}_0 - \alpha \int d^2 z \expval{T\Bar{T}(z) \Phi_1^R(x_1)... \Phi_N^R(x_N)}_{0} + \mathcal{O}(\alpha ^2)
\end{equation}

In order for the correlation function for $\Phi^R$ to be finite at $\epsilon \to 0$, one should choose $A(\epsilon) = \frac{2 \pi}{\epsilon} + \pi \log(\pi e ^{\gamma-1})$. Then in first order we get the answer
\begin{equation}
     \expval{\Phi_1^R(x_1)...\Phi_N^R(x_N)}_{def} = \biggl(1 + \alpha K\biggr)\expval{\Phi(x_1)...\Phi (x_N)}_0 +\mathcal{O}(\alpha^2),
\end{equation}
where $K$ is the differential operator
\begin{equation}
     K = \sum_{j,k\neq j}^{N}\biggl[ - \frac{\pi}{\bar{x}_j-
     \bar{x}_k}\bar{\Delta}_k \partial_{j} + \frac{\pi}{x_j-x_k}\Delta_j\bar{\partial}_k - 2\pi \log{|x_j- x_k|\partial_j \bar{\partial}_k}\biggr].
\end{equation}

\subsection{Second order correction}\label{sec:second_order}

In the second order of perturbation theory, a double integral arises
\begin{equation}\label{eq:second-order_int}
     \int d^2z \: d^2w \expval{T\bar{T}(z)T\bar{T}(w) \Phi_1(x_1)...\Phi_N(x_N)}_0
\end{equation}
We denote the insertion of the fields $\Phi_1(x_1)...\Phi_N(x_N)$ by $X$. Then, taking into account the Ward identities, we rewrite the integrand in the form (we omit the index indicating that the correlation functions were calculated in an unperturbed theory)
\begin{multline}
     \expval{T\bar{T}(z)T\bar{T}(w) X} = \frac{c^2}{4|z-w|^8}\expval{X} + \\
     + \frac{c}{2(z-w)^4}\biggl[\frac{2}{(\bar{z}-\bar{w})^2} + \frac{1}{\bar{z }-\bar{w}}\pdv{}{\bar{w}} + \sum_{i = 1}^{N}\left(\frac{\Bar{\Delta}_i}{(\bar{ z} - \bar{x}_i)^2} + \frac{1}{\bar{z} -\bar{x}_i}\pdv{\bar{x}_i} \right)\biggr]\expval{\bar{T}(\bar{w})X}+ h.c. +\\
     + \biggl[\frac{2}{(z-w)^2} + \frac{1}{z-w}\pdv{}{w} + \sum_{i = 1}^{N}\left(\frac{ \Delta_i}{(z - x_i)^2} + \frac{1}{z -x_i}\pdv{x_i} \right)\biggr]\cdot\\
     \cdot \biggl[\frac{2}{(\bar{z}-\bar{w})^2} + \frac{1}{\bar{z}-\bar{w}}\pdv{} {\bar{w}} + \sum_{j = 1}^{N}\left(\frac{\Bar{\Delta}_j}{(\bar{z} - \bar{x}_j)^2 } + \frac{1}{\bar{z} -\bar{x}_j}\pdv{\bar{x}_j} \right)\biggr]\expval{T(w)\bar{T}( \bar{w})X}
\end{multline}
The first term (we will call it the CC term) cancels with the contribution coming from the partition function, followed by the second term (CB term) and its complex conjugate partner (BC term), as well as the most complicated term (BB term).

One can do the following transformation under the integral sign
\begin{equation}
\begin{split}
    \int d^2w \: \frac{1}{z-w}\pdv{}{w}\:(...) &= \int d^2w \: \pdv{}{w}\left(\frac{ 1}{z-w}(...)\right) - \int d^2w \: \frac{1}{(z-w)^2} (...) 
    \\
    &= - \int d^2w \: \frac{ 1}{(z-w)^2} (...)
\end{split}
\end{equation}
Then the integrand in (\ref{eq:second-order_int}) will take the form
\begin{multline}
     \expval{T\bar{T}(z)T\bar{T}(w) X} = \frac{c^2}{4|z-w|^8}\expval{X} + \\
     + \frac{c}{2(z-w)^4}\biggl[\frac{1}{(\bar{z}-\bar{w})^2} + \sum_{i = 1}^{N }\left(\frac{\Bar{\Delta}_i}{(\bar{z} - \bar{x}_i)^2} + \frac{1}{\bar{z} -\bar{x }_i}\pdv{\bar{x}_i} \right)\biggr]\expval{\bar{T}(\bar{w})X}+ h.c. +\\
     + \biggl[\frac{1}{(z-w)^2} + \sum_{i = 1}^{N}\left(\frac{\Delta_i}{(z - x_i)^2} + \frac{ 1}{z -x_i}\pdv{x_i} \right)\biggr]
     \\
     \times \biggl[\frac{1}{(\bar{z}-\bar{w})^2} + \sum_{j = 1}^{N}\left(\frac{\Bar{\Delta }_j}{(\bar{z} - \bar{x}_j)^2} + \frac{1}{\bar{z} -\bar{x}_j}\pdv{\bar{x}_j } \right)\biggr]\expval{T(w)\bar{T}(\bar{w})X}
\end{multline}

\subsubsection{CB-term}

The CB-term is the integral that can be written as follows

\begin{multline}
     \int d^2z\:d^2w\:\frac{c}{2(z-w)^4}\biggl[\frac{1}{(\bar{z}-\bar{w})^2} + \sum_{i = 1}^{N}\left(\frac{\Bar{\Delta}_i}{(\bar{z} - \bar{x}_i)^2} + \frac{1} {\bar{z} -\bar{x}_i}\pdv{\bar{x}_i} \right)\biggr]\expval{\bar{T}(\bar{w})X} 
     \\
     = \int d^2z\:d^2w\:\frac{c}{2(z-w)^4}\sum_{i = 1}^{N}\frac{\Bar{\Delta}_i}{( \bar{z} - \bar{x}_i)^2} \sum_{j = 1}^{N}\left(\frac{\Bar{\Delta}_j}{(\bar{w} - \bar{x}_j)^2} + \frac{1}{\bar{w} -\bar{x}_j}\pdv{\bar{x}_j} \right)\expval{X}
     \\
     +\int d^2z\:d^2w\:\frac{c}{2(z-w)^4} \sum_{i = 1}^N \frac{1}{\bar{z} -\bar{ x}_i}\biggl(\frac{\bar{\Delta}_i}{(\bar{w}-\bar{x}_i)^3} + \frac{\bar{\Delta }_i+1}{(\bar{w}-\bar{x}_i)^2}\pdv{}{\bar{x}_i}
     \\
     +\frac{1}{\bar{w}-\bar{x}_i}\pdv[2]{}{\bar{x}_i}
     +\sum_{j\neq i}^{N}\frac{\Bar{\Delta}_j}{(\bar{w} - \bar{x}_j)^2}\pdv{}{\bar{ x}_i}+\sum_{j\neq i}^{N}\frac{1}{\bar{w} - \bar{x}_j}\pdv[]{}{\bar {x}_i}{\bar{x}_j} \biggr)\expval{X}
\end{multline}

Let us calculate the integrals over $z$. These integrals are convergent for $d = 2$, so they do not need to be regularized. These are the integrals $I_4^2$ and $I_4^1$, which are calculated in the Appendix \ref{appendix:intmn} (see also Table \ref{table:ints}).

Taking into account all of the above, as well as the properties of dimensional regularization, the CB term can be rewritten as
\begin{multline}
     -\int d^2w\:\frac{\pi c}{6} \sum_{i = 1}^N \frac{1}{(w-x_i)^3}\biggl(\frac{\bar{ \Delta}_i}{(\bar{w}-\bar{x}_i)^3}\expval{X} + \frac{\bar{\Delta}_i+1}{(\bar{w}- \bar{x}_i)^2}\pdv{}{\bar{x}_i}\expval{X}+
     \\
     +\frac{1}{\bar{w}-\bar{x}_i}\pdv[2]{}{\bar{x}_i}\expval{X}
     +\sum_{j\neq i}^{N}\frac{\Bar{\Delta}_j}{(\bar{w} - \bar{x}_j)^2}\pdv{}{\bar{ x}_i}\expval{X}+\sum_{j\neq i}^{N}\frac{1}{\bar{w} - \bar{x}_j}\pdv[]{}{\bar {x}_i}{\bar{x}_j}\expval{X} \biggr)
     \\
     = -\int d^2w\:\frac{\pi c}{6} \sum_{i = 1}^N \frac{1}{(w-x_i)^3}\biggl(\sum_{j\neq i}^{N}\frac{\Bar{\Delta}_j}{(\bar{w} - \bar{x}_j)^2}\pdv{}{\bar{x}_i}\expval {X}+\sum_{j\neq i}^{N}\frac{1}{\bar{w} - \bar{x}_j}\pdv[]{}{\bar{x}_i}{ \bar{x}_j}\expval{X} \biggr)
\end{multline}
To move further, we also need the integrals $I_2^3$ and $I_1^3$.

Collecting all terms together we get the final answer for the CB term
\begin{equation}
     \frac{\pi^2 c}{12}\sum_{i = 1, \: j \neq i}^{N} \frac{1}{(\bar{x}_i-\bar{x}_j )^2}\pdv[2]{}{\bar{x}_i}{\bar{x}_j}\expval{X}.
\end{equation}

\subsubsection{BB-term}
BB-term is defined as follows 
\begin{multline}\label{eq:bb-term}
    \int d^2w \:d^2z \: \biggl[\frac{1}{(z-w)^2} + \sum_{i = 1}^{N}\left(\frac{\Delta_i}{(z - x_i)^2} + \frac{1}{z -x_i}\pdv{x_i} \right)\biggr]\times
    \\
    \times \biggl[\frac{1}{(\bar{z}-\bar{w})^2} + \sum_{j = 1}^{N}\left(\frac{\Bar{\Delta}_j}{(\bar{z} - \bar{x}_j)^2} + \frac{1}{\bar{z} -\bar{x}_j}\pdv{\bar{x}_j} \right)\biggr]\expval{T(w)\bar{T}(\bar{w})X}.
\end{multline}

Here we have different integrals, some of them diverge and they should be regularized. Consider a simple 'factorized' part. 

\begin{equation}
    \text{Fact} = \int d^2w \:d^2z \:\sum_{i = 1}^{N}\left(\frac{\Delta_i}{(z - x_i)^2} + \frac{1}{z -x_i}\pdv{x_i} \right)\sum_{j = 1}^{N}\left(\frac{\Bar{\Delta}_j}{(\bar{z} - \bar{x}_j)^2} + \frac{1}{\bar{z} -\bar{x}_j}\pdv{\bar{x}_j} \right)\expval{T(w)\bar{T}(\bar{w})X}
\end{equation}
All its divergences are as follows (for details see appendix \ref{int:fact})
\begin{equation}
    \begin{split}
        \frac{2 \pi^2}{\epsilon}\sum_{i,k, j \neq i, l \neq k} \frac{\Delta_i}{x_i-x_j}\bar{\partial}_j\partial_l\bar{\partial}_k\expval{X} -\frac{2 \pi^2}{\epsilon}\sum_{i,k, j \neq i, l \neq k} \frac{\bar{\Delta}_i}{\bar{x}_i-\bar{x}_j}\partial_j\partial_l\bar{\partial}_k\expval{X}-
        \\
        -\frac{2\pi^2}{\epsilon}\sum_{i, j \neq i} \sum_{l, k \neq l} \left(\frac{\bar{\Delta}_k}{\bar{x}_l-\bar{x}_k}\partial_l - \frac{\Delta_l}{x_l - x_k}\bar{\partial}_k - 4 \log(m|x_l - x_k|)\partial_l\bar{\partial}_k\right)\partial_i\bar{\partial}_j\expval{X}
        \\
         +\left(\frac{2\pi}{\epsilon}\right)^2\sum_{i; j \neq i;l \neq i,k; k \neq j}^N \partial_i \bar{\partial}_j \partial_l \bar{\partial}_k\expval{X}
    \end{split}
\end{equation}
There can also be divergences from the unfactorized part:
\begin{multline}
    \int d^2w \:d^2z \: \biggl[\frac{1}{(z-w)^2}\sum_{i = 1}^{N}\left(\frac{\Bar{\Delta}_i}{(\bar{z} - \bar{x}_i)^2} + \frac{1}{\bar{z} -\bar{x}_i}\pdv{\bar{x}_i} \right) +\sum_{i = 1}^{N}\left(\frac{\Delta_i}{(z - x_i)^2} + \frac{1}{z -x_i}\pdv{x_i} \right)\frac{1}{(\bar{z}-\bar{w})^2}\biggr]
    \\
    \times \sum_{l = 1}^{N}\left(\frac{\Delta_l}{(w - x_l)^2} + \frac{1}{w -x_l}\pdv{x_l} \right) \sum_{k = 1}^{N}\left(\frac{\Bar{\Delta}_k}{(\bar{w} - \bar{x}_k)^2} + \frac{1}{\bar{w} -\bar{x}_k}\pdv{\bar{x}_k} \right)\expval{X}
\end{multline}
The integrals that arise here, which we need to calculate, are as follows:
\begin{equation}\label{eq:ugly_int}
    I_{ac}^{bd} (x, y, t) = \int d^2z \: d^2w \frac{1}{(z-w)^a}\frac{1}{(\bar{z}- \bar{x})^b} \frac{1}{(w-y)^c} \frac{1}{(\bar{w}-\bar{t})^d},
\end{equation}
where $a, \:b, \:c, \:d$ take values 1 and 2, and $a$ and $b$ cannot both be equal to 1 at the same time. One can see that the case when $y = x$ or $t = x$ does not lead to new integrals (they can be calculated by the first-order methods). One can note that for a two-point function ($N = 2$), integrals with all different points $x, \: y$ and $t$ do not arise. We leave the analysis of the divergent part of the integral (\ref{eq:ugly_int}) for a future publication. Here we consider only the case of two-point function.

\subsection{Second order correction for the case of a two-point function}

Let us introduce the notation:
\begin{equation}
    \partial = \partial_x = \pdv{}{x}, \: \partial_y = \pdv{}{y}, \: \Delta = \Delta_1 = \Delta_2, \: \bar{\Delta} = \bar{\Delta}_1 = \bar{\Delta}_2.
\end{equation}
Also in the case of two-point function we have
\begin{equation}
    \partial_x \expval{\Phi(x) \Phi(y)}_0 + \partial_y \expval{\Phi(x) \Phi(y)} = 0.
\end{equation}
Let us write down the divergences in the case of a two-point function. Second order pole:
\begin{equation}
    \left(\frac{2\pi}{\epsilon}\right)^2(\partial_x^2\bar{\partial}_y^2 + 2 \partial_x\partial_y\bar{\partial}_x\bar{\partial}_y +   \partial_y^2\bar{\partial}_x^2) \expval{X} = \left(\frac{2\pi}{\epsilon}\right)^2(\partial_x\bar{\partial}_y + \partial_y\bar{\partial}_x)^2 \expval{X} = 4\left(\frac{2\pi}{\epsilon}\right)^2\partial^2\bar{\partial}^2 \expval{X}
\end{equation}
We write down here all the poles in the factorized part of the BB-term
\begin{multline}
    4\frac{(2\pi)^2}{\epsilon} \log(m|x-y|) \partial^2\bar{\partial}^2 \expval{X} + 4\frac{(2\pi)^2}{\epsilon}\frac{\bar{\Delta}}{\bar{y}-\bar{x}}\partial^2\bar{\partial}\expval{X} + 4 \frac{(2\pi)^2}{\epsilon}\frac{\Delta}{y-x}\partial\bar{\partial}^2\expval{X} -
    \\
    - 2 \frac{(2\pi)^2}{\epsilon} \frac{\Delta}{(y-x)^2}\bar{\partial}^2 - 2 \frac{(2\pi)^2}{\epsilon} \frac{\bar{\Delta}}{(\bar{y}-\bar{x})^2}\partial + 4\left(\frac{2\pi}{\epsilon}\right)^2\partial^2\bar{\partial}^2 \expval{X}.
\end{multline}

Let us consider the unfactorized part of the BB-term. Essentially we need to consider the following integral
\begin{multline}
    \int d^2w\:d^2z \frac{1}{(z-w)^2} \left(\frac{\bar{\Delta}}{(\bar{z} - \bar{x})^2} + \frac{\bar{\Delta}}{(\bar{z} - \bar{y})^2} + \frac{1}{\bar{z} - \bar{x}}\pdv{}{\bar{x}} + \frac{1}{\bar{z} - \bar{y}}\pdv{}{\bar{y}}\right)
    \\
    \times\left(\frac{\Delta}{(w - x)^2} + \frac{\Delta}{(w - y)^2} + \frac{1}{w - x}\pdv{}{x} + \frac{1}{w - y}\pdv{}{y}\right)
    \\
    \times\left(\frac{\bar{\Delta}}{(\bar{w} - \bar{x})^2} + \frac{\bar{\Delta}}{(\bar{w} - \bar{y})^2} + \frac{1}{\bar{w} - \bar{x}}\pdv{}{\bar{x}} + \frac{1}{\bar{w} - \bar{y}}\pdv{}{\bar{y}}\right)\expval{X}.
\end{multline}
We introduce the notations of differential operators that constitutes the required integral.
\begin{align}
    D(w|x,y) &= \frac{\Delta}{(w - x)^2} + \frac{\Delta}{(w - y)^2} + \frac{1}{w - x}\pdv{}{x} + \frac{1}{w - y}\pdv{}{y} 
    \\
    \bar{D}(\bar{w}|\bar{x},\bar{y}) &= \frac{\bar{\Delta}}{(\bar{w} - \bar{x})^2} + \frac{\bar{\Delta}}{(\bar{w} - \bar{y})^2} + \frac{1}{\bar{w} - \bar{x}}\pdv{}{\bar{x}} + \frac{1}{\bar{w} - \bar{y}}\pdv{}{\bar{y}}
    \\
    \bar{D}_{\bar{x}}(\bar{w}|\bar{x},\bar{y}) &=\frac{2\bar{\Delta}}{(\bar{w} - \bar{x})^3} +  \frac{\bar{\Delta} + 1}{(\bar{w} - \bar{x})^2} \pdv{}{\bar{x}}+ \frac{1}{\bar{w} - \bar{x}}\pdv[2]{}{\bar{x}} + \frac{\bar{\Delta}}{(\bar{w} - \bar{y})^2} \pdv{}{x}+ \frac{1}{\bar{w} - \bar{y}}\pdv{}{x}\pdv{}{\bar{y}}
\end{align}
Then the integral can be divided into three parts and take the form:
\begin{equation}
    I_1 + I_x + I_y,
\end{equation}
where we introduced the notation
\begin{align}
    I_1 &= \int d^2w\:d^2z \frac{1}{(z-w)^2} \left(\frac{\bar{\Delta}}{(\bar{z} - \bar{x})^2} + \frac{\bar{\Delta}}{(\bar{z} - \bar{y})^2}\right) D(w|x,y) \bar{D}(\bar{w}|\bar{x},\bar{y})\expval{X}, \\
    I_x &= \int d^2w\:d^2z \frac{1}{(z-w)^2} \frac{1}{\bar{z} - \bar{x}} D(w|x,y) \bar{D}_{\bar{x}}(\bar{w}|\bar{x},\bar{y})\expval{X},\\
    I_y &=\int d^2w\:d^2z \frac{1}{(z-w)^2} \frac{1}{\bar{z} - \bar{x}} D(w|x,y) \bar{D}_{\bar{y}}(\bar{w}|\bar{x},\bar{y})\expval{X}.
\end{align}
The main idea is to take integrals directly in $d = 2$, and if they lead to divergences, substitute the integrals by their expansion at $d \to 2$. Acting according to this scheme, we conclude that $I_1$ can lead only to terms finite in $d\to2$ limit. That can be easily seen as far as integral over $z$ is of the form $I_2^2$, which behaves like $\mathcal{O}(d-2)$, while the integral over $w$ cannot produce divergences higher than $\frac{1}{d-2}$.

Let us move further and calculate the $I_x$ integral
\begin{multline}
    \int d^2w\:d^2z \frac{1}{(z-w)^2} \frac{1}{\bar{z} - \bar{x}} D(w|x,y) \bar{D}_{\bar{x}}(\bar{w}|\bar{x},\bar{y})\expval{X} \sim
    \\
    \sim \int d^2w \: \frac{1}{w-x} \left( \frac{\Delta}{(w - x)^2} + \frac{\Delta}{(w - y)^2} + \frac{1}{w - x}\pdv{}{x} + \frac{1}{w - y}\pdv{}{y}\right)
    \\
    \times \left(\frac{2\bar{\Delta}}{(\bar{w} - \bar{x})^3} +  \frac{\bar{\Delta} + 1}{(\bar{w} - \bar{x})^2} \pdv{}{\bar{x}}+ \frac{1}{\bar{w} - \bar{x}}\pdv[2]{}{\bar{x}} + \frac{\bar{\Delta}}{(\bar{w} - \bar{y})^2} \pdv{}{x}+ \frac{1}{\bar{w} - \bar{y}}\pdv{}{x}\pdv{}{\bar{y}}\right)
\end{multline}
When integrating over $w$, the divergences give only the terms with different external points of the form
\begin{equation}
    \frac{1}{(w-x)(\bar{w} - \bar{y})},
\end{equation}
and one can note that both $w$ and $\bar{w}$ should have exactly degree $-1$. Thus, of the terms $D(w|x,y)$, only the 2nd and 4th ones can potentially give divergences. Further, from $\bar{D}_{\bar{x}}(\bar{w}|\bar{x},\bar{y})$ only the 3rd and 5th terms can give divergences. Leaving only potentially divergent parts, we write
\begin{equation}
   -\pi \int d^2w \: \frac{1}{w-x} \left( \frac{\Delta}{(w - y)^2} + \frac{1}{w - y}\pdv{}{y}\right)
    \cdot \left(\frac{1}{\bar{w} - \bar{x}}\pdv[2]{}{\bar{x}} + \frac{1}{\bar{w} - \bar{y}}\pdv{}{x}\pdv{}{\bar{y}}\right)\expval{X}
\end{equation}
To integrate one needs to decompose the holomorphic part into simple terms using the formulas
\begin{align}
    \frac{1}{(w-x)(w-y)} &= \frac{1}{(x-y)(w-x)} - \frac{1}{(x-y)(w-y)}
    \\
    \frac{1}{(w-x)(w-y)^2} &= \frac{1}{(x-y)^2(w-x)} - \frac{1}{(x-y)^2(w-y)} - \frac{1}{(x-y)(w-y)^2}
\end{align}
and leave only the terms of degrees $-1$ both in $w$ and in $\bar{w}$, but with different points. We have
\begin{multline}
   -\pi \int d^2w \: \biggl( \frac{\Delta}{(x-y)^2(w-x)} - \frac{\Delta}{(x-y)^2(w-y)} - \frac{\Delta}{(x-y)(w-y)^2} +
    \\
    + \frac{1}{(x-y)(w-x)}\pdv{}{y} - \frac{1}{(x-y)(w-y)}\pdv{}{y}\biggr)
    \\
    \times \left(\frac{1}{\bar{w} - \bar{x}}\pdv[2]{}{\bar{x}} + \frac{1}{\bar{w} - \bar{y}}\pdv{}{x}\pdv{}{\bar{y}}\right)\expval{X}
\end{multline}
One can obtain the result
\begin{equation}
    \left(-\frac{(2\pi)^2}{\epsilon}\frac{\Delta}{(x-y)^2}\pdv{}{\bar{x}^2} + \frac{(2\pi)^2}{\epsilon(x-y)}\pdv{}{\bar{x}^2}\pdv{}{x}\right)\expval{X}
\end{equation}
Completely analogous procedure for $I_y$
leads to an answer
\begin{equation}
    \left(-\frac{(2\pi)^2}{\epsilon}\frac{\Delta}{(x-y)^2}\pdv{}{\bar{x}^2} - \frac{(2\pi)^2}{\epsilon(x-y)}\pdv{}{\bar{x}^2}\pdv{}{x}\right) \expval{X}
\end{equation}
The total divergent contributions of $I_x$ and $I_y$ are equal
\begin{equation}\label{eq:not_conj_res}
    -2\frac{(2\pi)^2}{\epsilon}\frac{\Delta}{(x-y)^2}\pdv{}{\bar{x}^2}\expval{X}
\end{equation}
A similar contribution will be complex conjugate to this one. It comes from the term in the original integral
\begin{multline}
    \int d^2w\:d^2z  \left(\frac{\Delta}{(z - x)^2} + \frac{\Delta}{(z - y)^2} + \frac{1}{z - x}\pdv{}{x} + \frac{1}{z - y}\pdv{}{y}\right)\frac{1}{(\bar{z}-\bar{w})^2}
    \\
    \times\left(\frac{\Delta}{(w - x)^2} + \frac{\Delta}{(w - y)^2} + \frac{1}{w - x}\pdv{}{x} + \frac{1}{w - y}\pdv{}{y}\right)
    \\
    \times\left(\frac{\bar{\Delta}}{(\bar{w} - \bar{x})^2} + \frac{\bar{\Delta}}{(\bar{w} - \bar{y})^2} + \frac{1}{\bar{w} - \bar{x}}\pdv{}{\bar{x}} + \frac{1}{\bar{w} - \bar{y}}\pdv{}{\bar{y}}\right)\expval{X}
\end{multline}
This contribution will simply be the complex conjugate of (\ref{eq:not_conj_res})
\begin{equation}
    -2\frac{(2\pi)^2}{\epsilon}\frac{\bar{\Delta}}{(\bar{x}-\bar{y})^2}\pdv{}{x^2}\expval{X}.
\end{equation}

\subsubsection{Renormalization}
In order to remove all divergences, it is necessary to renormalize the fields. We will look for it in the form
\begin{multline}
    \Phi^R(x) = \Phi(x) + \alpha A(\epsilon) L_{-1}\bar{L}_{-1} \Phi(x) + \alpha^2 B_{11}(\epsilon) L_{-1}^2\bar{L}_{-1}^2 \Phi(x) +\alpha^2 B_{21}(\epsilon) L_{-2}\bar{L}_{-1}^2 \Phi(x)
    +
    \\
    + \alpha^2 B_{12}(\epsilon) L_{-1}^2\bar{L}_{-2} \Phi(x) + \alpha^2 B_{22}(\epsilon) L_{-2}\bar{L}_{-2}\Phi(x) + \mathcal{O}(\alpha^3)
\end{multline}
In fact, since we consider a two-point correlation function, and also since the correlation functions of primary fields with quasi-primary ones are equal to zero, then it is correct to look for renormalizations among the descendants of the form
\begin{equation}
    D_{-1}^2\Phi, \: D_{-2}\Phi,
\end{equation}
where 
\begin{align}
    D_{-1} &= L_{-1},
    \\
    D_{-2} &= L_{-2} + \mu L_{-1}^2, \: \mu = -\frac{3}{2(2\Delta + 1)}
\end{align}
Thus, $D_{-2}\Phi$ is the quasi-primary field. One can expand the renormalization to the second order using two bases:
\begin{multline}
    B_{11}L_{-1}^2\bar{L}_{-1}^2 + B_{21}L_{-2}\bar{L}_{-1}^2 + B_{12}L_{-1}^2\bar{L}_{-2} + B_{22}L_{-2}\bar{L}_{-2} =
    \\
    =\Tilde{B}_{11}D_{-1}^2\bar{D}_{-1}^2 + \Tilde{B}_{21}D_{-2}\bar{D}_{-1}^2 + \Tilde{B}_{12}D_{-1}^2\bar{D}_{-2} + \Tilde{B}_{22}D_{-2}\bar{D}_{-2}
\end{multline}
Correspondence of the expansion coefficients is given by
\begin{align}
    \Tilde{B}_{11} &= B_{11} - \mu B_{12} - \mu B_{21} + \mu^2 B_{22}
    \\
    \Tilde{B}_{12} &= B_{12} - \mu B_{22}
    \\
    \Tilde{B}_{21} &= B_{21} - \mu B_{22}
    \\
    \Tilde{B}_{22} &= B_{22}
\end{align}
In this case, the renormalization will look this way
\begin{equation}
    \Phi^{R} (x) = \Phi(x) + \alpha A(\epsilon)L_{-1}\bar{L}_{-1}\Phi(x) +  \alpha^2\Tilde{B}_{11} D_{-1}^2\bar{D}_{-1}^2\Phi(x)
\end{equation}
Other terms, containing $D_{-2}$ or $\bar{D}_{-2}$ cannot be determined considering the two-point correlation function.

Let us calculate the correlation function of such fields in the perturbed theory. Let's remember that
\begin{equation}
    \expval{L_{-2} \Phi(x) \Phi(y)}_0 = \expval{ \Phi(x) L_{-2}\Phi(y)}_0 = \left(\frac{\Delta}{(x-y)^2} - \frac{1}{x-y}\partial_x\right)\expval{\Phi(x) \Phi(y)}_0
\end{equation}
\begin{equation}
    \expval{\bar{L}_{-2} \Phi(x) \Phi(y)}_0 = \expval{ \Phi(x) \bar{L}_{-2} \Phi(y)}_0 = \left(\frac{\Delta}{(\bar{x}-\bar{y})^2} - \frac{1}{\bar{x}-\bar{y}}\bar{\partial}_x\right)\expval{\Phi(x) \Phi(y)}_0
\end{equation}
and use this when calculating the correlation functions:
\begin{multline}
    \expval{\Phi^R(x_1)\Phi^R(x_2)}_{def} = \expval{\Phi^R(x_1)\Phi^R(x_2)}_{0} - \alpha \int d^2z \: \expval{T\bar{T}(z)\Phi^R(x_1)\Phi^R(x_2)}_{0}
    \\
    +\frac{\alpha^2}{2}\int d^2z\:d^2w\: \biggl(\expval{T\bar{T}(z)T\bar{T}(w)\Phi^R(x_1)\Phi^R(x_2)}_{0} - \expval{T\bar{T}(z)T\bar{T}(w)}_0\expval{\Phi^R(x_1)\Phi^R(x_2)}_0\biggr) + \mathcal{O}(\alpha^2)
    \\
    = \expval{X}_{0} + 2\alpha A(\epsilon)\partial \bar{\partial}\expval{X}_{0} + \alpha^2 A^2(\epsilon)(\partial \bar{\partial})^2\expval{X}_{0} + 2 \alpha^2 \Tilde{B}_{11}(\epsilon) (\partial\bar{\partial})^2 \expval{X}_0 
    \\
    - \alpha \int d^2z \expval{T\bar{T}(z)X}_0 - 2\alpha^2 A(\epsilon) \int d^2 z \: \expval{T\bar{T}(z)\partial \bar{\partial}\Phi(x_1)\Phi(x_2)}_0 
    \\
    + \frac{\alpha^2}{2}\int d^2z\:d^2w\: \biggl(\expval{T\bar{T}(z)T\bar{T}(w)\Phi(x_1)\Phi(x_2)}_{0} - \expval{T\bar{T}(z)T\bar{T}(w)}_0\expval{\Phi(x_1)\Phi(x_2)}_0\biggr) + \mathcal{O}(\alpha^2)
\end{multline}
Now the problem is as follows: select such coefficient $\Tilde{B}_{11}(\epsilon)$, that in the final expression for $\expval{\Phi^R(x_1)\Phi^R(x_2)}_{def}$ all the divergences are cancelled. The divergences at the order $\alpha$ were mutually cancelled by the choice
\begin{equation}
    A(\epsilon) = \frac{2\pi}{\epsilon} + \pi \log{m^2}
\end{equation}
In order to find out the divergences at $\alpha^2$ let us calculate the integral
\begin{multline}
    \int d^2 z \: \expval{T\bar{T}(z)\partial \bar{\partial} \Phi(x_1)\Phi(x_2)}_0 =
    \\
    = 2\pdv{}{x_1} \pdv{}{\bar{x}_1}\left[\biggl(\frac{\pi}{\bar{x}_1-
    \bar{x}_2}\bar{\Delta} \partial + \frac{\pi}{x_1-x_2}\Delta\bar{\partial} + \biggl(\frac{2\pi}{\epsilon}+ 2\pi \log{(|x_1-x_2|m)}\biggr)\partial \bar{\partial} \biggr) \expval{X}_0\right] =
    \\
    = 2\pi \biggl(\frac{2}{\epsilon}\partial^2\bar{\partial}^2 + 2 \log(m|x_1-x_2|)\partial^2\bar{\partial}^2 + \frac{\Delta+1}{x_1-x_2}\partial\bar{\partial}^2 + \frac{\bar{\Delta} + 1}{\bar{x}_1-\bar{x}_2}\partial^2\bar{\partial} -
    \\
    - \frac{\Delta}{(x_1-x_2)^2}\bar{\partial}^2 - \frac{\bar{\Delta}}{(\bar{x}_1-\bar{x}_2)^2}\partial^2\biggr)\expval{X}_0
\end{multline}
A remarkable fact is that divergences of the form
\begin{equation}
    4\frac{(2\pi)^2}{\epsilon} \log(m|x_1-x_2|) \partial^2\bar{\partial}^2 \expval{X}
\end{equation}
are cancelled completely.

In the following let us remind that
\begin{equation}
    \expval{\Phi(x_1)\Phi(x_2)}_0 = \frac{C}{|x_1-x_2|^{2\Delta + 2\bar{\Delta}}},
\end{equation}
and hence follows the relations
\begin{align}
    \partial  \expval{\Phi(x_1)\Phi(x_2)}_0 &= (-2\Delta) \cdot \expval{\Phi(x_1)\Phi(x_2)}_0,
    \\
    \partial^2  \expval{\Phi(x_1)\Phi(x_2)}_0 &= (-2\Delta)\cdot (-2\Delta - 1)\cdot \expval{\Phi(x_1)\Phi(x_2)}_0,
    \\
    \left(\frac{\Delta}{(x_1-x_2)^2} - \frac{1}{x_1-x_2}\partial_x\right) \expval{\Phi(x_1)\Phi(x_2)}_0 &= 3\Delta \cdot \expval{\Phi(x_1)\Phi(x_2)}_0.
\end{align}
Then, matching the coefficients at different poles and different powers of $\Delta$ and $\bar{\Delta}$, we obtain that
\begin{equation}
    \Tilde{B}_{11} = \frac{1}{2} A(\epsilon)^2 +  \frac{\pi^2}{\epsilon}\left(\frac{1}{2\Delta+1} + \frac{1}{2\bar{\Delta} + 1}\right)
\end{equation}
The term $\frac{1}{2} A(\epsilon)^2$ is a second-order pole and is consistent with that obtained in \cite{Cardy_2019}.

\section{Conclusion and discussion}

In this study, we have computed the effect of $T\bar{T}$-deformation on correlation functions of primary fields in conformal field theory. We provide the full calculation of first order correction in perturbation theory. In first order, we see that primary fields are mixed with their Laplacians in the process of the renormalization.

In the second order we succeeded to calculate the case of two-point functions. We see the highest divergence of order $\mathcal{O}(\epsilon^{-2})$, which is exactly $\frac{1}{2} A(\epsilon)^2$, is the term coming from the renormalization term
\begin{equation}
    \Phi^{R} (x) = e^{\alpha A(\epsilon) \partial \bar{\partial}} \Phi(x).
\end{equation}
Note that due to the difference in renormalization scheme in order to compare this expression with the one given in \cite{Cardy_2019} one needs to replace the divergence $\frac{1}{\epsilon}$ with $\log \varepsilon$, where $\varepsilon$ -- is the excluded hard-disc radius. We have also succeeded in deriving other than this "exponential" renormalization terms, that are less divergent and are of order $\mathcal{O}(\epsilon^{-1})$. One can note that this correction is not universal for all fields but essentially depends on conformal and anti-conformal dimensions.

\acknowledgments

The author is grateful to Alexander Zamolodchikov and Alexei Litvinov for valuable discussions and advices on the manuscript. The work was done within the framework of the state assignment: 1.1.3-0023/24.

\paragraph{Competing interests}
\leavevmode 
\medskip

The author declares no competing interests.

\newpage

\appendix
\label{appendix}
\section{First order integrals}
Useful formulas for calculations:
\begin{equation}
\begin{split}
     \int d^dV \frac{(V^2)^a}{(V^2 + D)^b} &= \pi^{\frac{d}{2}}\frac{\Gamma(b-a-d/2 )\Gamma(a+d/2)}{\Gamma(b)\Gamma(d/2)}\cdot \frac{1}{D^{b-a-d/2}}
    \\
     \frac{1}{A^a B^b} &= \frac{\Gamma(a+b)}{\Gamma(a)\Gamma(b)}\int_0^1du \frac{u^{a-1 }(1-u)^{b-1}}{(Au+B(1-u))^{a+b}}
\end{split}
\end{equation}

\subsection{General m, n} \label{appendix:intmn}
\begin{multline}
     I_m^n(x, y) = \int d^2z \frac{1}{(z-x)^m(\bar{z} - \bar{y})^n} = \int d^2z \frac{(\bar{z} - \bar{x})^m(z-y)^n}{|z-x|^{2m}|z-y|^{2n}} =
     \\
     = \lim_{d \to 2} \int d^dz \frac{((Z - X)\nu_+)^m((Z-Y)\nu_-)^n}{|Z-X|^{2m}|Z-Y|^{2n}},
\end{multline}
where $\nu_+ = (1, -i, 0, 0, ...)$ and $\nu_- = (1, i, 0, 0, ...)$ are $d$-dimensional projectors, and $Z$, $X$, $Y$ are some $d$-dimensional vectors, such that $Z\cdot\nu_+ = \Bar{z}$, $Z \cdot \nu_- = z$ etc. Let us make the substitution $W = Z - Y$, $A = X - Y$. Then the integral will take the form
\begin{equation}
     \int d^dW \frac{((W-A)\nu_+)^m(W\nu_-)^n}{|W-A|^{2m}|W|^{2n}}.
\end{equation}
Using the Feynman parameter and formula
\begin{equation}\label{eq:int_W}
    \frac{1}{B^a C^b} = \frac{\Gamma(a+b)}{\Gamma(a)\Gamma(b)}\int_0^1du \frac{u^{a-1 }(1-u)^{b-1}}{(Bu+C(1-u))^{a+b}}
\end{equation}
with $B =|W-A|^2$, $C = |W|^2$ the denominator in (\ref{eq:int_W}) can be rewritten as
\begin{equation}
     \frac{1}{(|W-A|^2)^m(|W|^2)^n} = \int_0^1 du \frac{u^m(1-u)^n}{((W-Au)^2+u(1-u)A^2)^{m+n} }.
\end{equation}
Next, one can make the replacement $V = W - uA$ and $D = A^2u(1-u)$. As a result, the integral takes the form
\begin{equation}\label{int_unmod}
     \int_0^1 du\: u^m(1-u)^n \int d^dV \frac{((V-A(1-u))\nu_+)^m((V + uA)\nu_-)^n}{(V^2+ D)^{m+n}}.
\end{equation}

Let us work with the numerator
\begin{equation}\label{VV_mult}
    ((V-A(1-u))\nu_+)^m((V + uA)\nu_-)^n = \prod_{i=1}^m\prod_{j=1}^n\biggl(V^{\alpha_i} - A^{\alpha_i} (1-u)\biggr)\biggl(V^{\beta_j} + u A^{\beta_j}\biggl) \nu_{+\alpha_i}\nu_{-\beta_j}.
\end{equation}
Since this term is under the integral over $V$, the denominator of the integrand is invariant under the change $V \to -V$, then only the terms with an even number of $V^{\alpha}$ will give a non-zero contribution to the integral. One can expand the brackets in (\ref{VV_mult}) to obtain
\begin{equation}\label{mult_expansion}
    \begin{split}
        ((V-A(1-u))\nu_+)^m((V + uA)\nu_-)^n &= \sum_{k=0}^{m}\sum_{\substack{l=0,\:\\l+k \in 2 \mathbb{Z}}}^n (-1)^{m-k}C_m^k C_n^l u^{n-l}(1-u)^{m-k}
        \\
        \cdot V^{\alpha_1}\ldots  V^{\alpha_k}V^{\beta_1}\ldots V^{\beta_l}A^{\alpha_{k+1}}\ldots  A^{\alpha_{m}}&A^{\beta_{l+1}}\ldots A^{\beta_n}\nu_{+\alpha_1}\ldots\nu_{+\alpha_m}\nu_{-\beta_1}\ldots \nu_{-\beta_n},
     \end{split}
\end{equation}
where the emerging scalar product of the vectors should be treated as follows:
\begin{equation}
     A^\beta\nu_{-\beta} = x-y, \: A^{\alpha}\nu_{+\alpha} = \bar{x} - \bar{y}, \: \nu_+^{\alpha}\nu_{+\alpha} = 0, \: \nu_-^{\alpha}\nu_{-\alpha} = 0, \: \nu_+^{\alpha}\nu_{-\alpha} = 2.
\end{equation}

Let us use the averaging rule, according to which we can make the substitution under the integral sign
\begin{equation}\label{averaging_rule}
     V^{\alpha_1} \ldots V^{\alpha_{2N}} \to \frac{V^{2N}}{P_N(d)}\sum_{(r_1,\ldots,r_N)\in \mathcal{P}_2(\{\alpha_1,\ldots,\alpha_{2N}\})} g^{r_i(1) r_i(2)},
\end{equation}
where $g^{\alpha \beta}$ is the eucledian metric; we denoted the set of all partitions of the set $A$ into pairs by $\mathcal{P}_2(A)$, $r_i(1)$ and $r_i(2)$ are the first and the second elements of the i'th pair respectively and $P_N(d)$ is a normalizing factor that is equal to
\begin{equation}
    P_N(d) = \prod_{i=1}^N (d+2(i-1)) = 2^N \frac{\Gamma(d/2+N-1)}{\Gamma(d/2)}.
\end{equation}
In the article we are faced with the cases $N = 1, 2$, that explicitly are written as
\begin{align}
    V^{\alpha} V^{\gamma} &\to \frac{V^2}{d} g^{\alpha \gamma},\\
     V^\alpha V^\beta V^\gamma V^\sigma &\to \frac{V^4}{d(d+2)}(g^{\alpha\beta}g^{\gamma\sigma } + g^{\gamma\beta}g^{\alpha \sigma} + g^{\sigma\beta}g^{\alpha \gamma}).
\end{align}
One can now move further and calculate the integral over $V$ by using the formula
\begin{equation}\label{integration_V}
    \int d^dV \frac{(V^2)^a}{(V^2 + D)^b} = \pi^{\frac{d}{2}}\frac{\Gamma(b-a-d/2 )\Gamma(a+d/2)}{\Gamma(b)\Gamma(d/2)}\cdot \frac{1}{D^{b-a-d/2}}
\end{equation}
and the integral over $u$ by using the formula
\begin{equation}\label{integration_u}
    \int_0^1 du \: u^s (1-u)^t = \frac{\Gamma(1+s)\Gamma(1+t)}{\Gamma(2+s+t)}.
\end{equation}
After substitution (\ref{mult_expansion}) in (\ref{int_unmod}) and integrating over $V$ and $u$ using (\ref{integration_V}) and (\ref{integration_u}) one can obtain the result
\begin{equation}
    \begin{split}         &\sum_{k=0}^{m}\sum_{\substack{l=0,\:\\l+k \in 2 \mathbb{Z}}}^n (-1)^{m-k}\frac{C_m^k C_n^l}{2^N} \pi^{\frac{d}{2}}
    \\
    &\times \frac{\Gamma(m+n-k-d/2)\Gamma(k+l+d/2)\Gamma(1+m+n-l)\Gamma(1+m+n-k)\Gamma(d/2)}{\Gamma(m+n)\Gamma(d/2)\Gamma(2+m+n-l-k)\Gamma(k+l-1+d/2)}
    \\
    &\times \sum_{(r_1,\ldots,r_N)\in \mathcal{P}_2(\{\alpha_1,\ldots,\alpha_{2N}\})} g^{r_i(1) r_i(2)} A^{\alpha_{k+1}}\ldots  A^{\alpha_{m}}A^{\beta_{l+1}}\ldots A^{\beta_n}\nu_{+\alpha_1}\ldots\nu_{+\alpha_m}\nu_{-\beta_1}\ldots \nu_{-\beta_n}
    \end{split}
\end{equation}

\subsection{m = 1, n = 1}
Let us look how does that work in particular case $m =1$ and $n=1$. We have to calculate the integral
\begin{multline}
     I_1^1(x, y) = \int d^2z \frac{1}{(z-x)(\bar{z} - \bar{y})} = \int d^2z \frac{(\bar{ z} - \bar{x})(z-y)}{|z-x|^2|z-y|^2} =
     \\
     = \lim_{d \to 2} \int d^dz \frac{((Z - X)\nu_+)((Z-Y)\nu_-)}{|Z-X|^2|Z-Y|^2},
\end{multline}
where $\nu_+ = (1, -i, 0, 0, ...)$ and $\nu_- = (1, i, 0, 0, ...)$ are $d$-dimensional projectors, and $Z$, $X$, $Y$ are $d$-dimensional vectors. Let's make the replacement $W = Z - Y$, $A = X - Y$. Then the integral will take the form
\begin{equation}
     \int d^dW \frac{((W-A)\nu_+)(W\nu_-)}{|W-A|^2|W|^2}.
\end{equation}
Using the Feynman parameter, the denominator can be reduced to the form
\begin{equation}
     \frac{1}{|W-A|^2|W|^2} = \int_0^1 du \frac{1}{((W-Au)^2+u(1-u)A^2)^2 }
\end{equation}
Next, you can enter $V = W - uA$ and $D = A^2u(1-u)$, and the integral takes the form
\begin{equation}
     \int_0^1 du \int d^dV \frac{((V-A(1-u))\nu_+)((V + uA)\nu_-)}{(V^2+ D)^2}
\end{equation}
Let us work with the numerator
\begin{equation}
     ((V-A(1-u))\nu_+)((V + uA)\nu_-) = \biggl(V^\alpha - A^\alpha (1-u)\biggr)\biggl(V^\beta + u A^\beta\biggl) \nu_{+\alpha}\nu_{-\beta}
\end{equation}
Since this term is under the integral over $V$, the denominator of the integrand is invariant under the change $V \to -V$, then only terms with an even number $V^{\alpha}$ need to be retained.
\begin{equation}
     \biggl(V^\alpha V^\beta - u(1-u) A^\alpha A^\beta\biggr)\nu_{+\alpha}\nu_{-\beta}
\end{equation}
Let us use the averaging rule, according to which we can make the substitution under the integral sign
\begin{equation}
     V^{\alpha} V^{\gamma} \to \frac{V^2}{d} g^{\alpha \gamma}
\end{equation}
Now the numerator is equal
\begin{equation}
     \frac{V^2}{d}(\nu_{+\alpha} \nu_-^{\alpha}) -u(1-u) (A^\alpha\nu_{+\alpha})(A^ \beta\nu_{-\beta}),
\end{equation}
Where
\begin{equation}
     A^\beta\nu_{-\beta} = x-y, \: A^{\alpha}\nu_{+\alpha} = \bar{x} - \bar{y}, \: \nu_+^{\alpha}\nu_{+\alpha} = 0, \: \nu_-^{\alpha}\nu_{-\alpha} = 0, \: \nu_+^{\alpha}\nu_{-\alpha} = 2.
\end{equation}
Emerging integrals:
\begin{align}
     &\int d^dV \frac{V^2}{(V^2+D)^2} = \pi^{d/2}\frac{\Gamma(1-d/2)\Gamma(1+ d/2)}{\Gamma(d/2)} D^{-1+d/2}
     \\
     &\int d^dV \frac{1}{(V^2+D)^2} = \pi^{d/2}\Gamma(2-d/2) D^{-2+d/2}
\end{align}
Substituting everything, we get the answer
\begin{align}
     \int d^dz \frac{1}{(z-x)(\bar{z} - \bar{y})} = \pi^{\frac{d}{2}}\int_0^1 du\biggl( \Gamma(1-d/2) - \Gamma(2-d/2)\biggr)\frac{1}{D^{1-\frac{d}{2}}} =
     \\
     =\frac{\pi^{\frac{d}{2}}\Gamma(1-d/2)\frac{d}{2}}{|x-y|^{2-d}}\int_0^1 du\:u^{\frac{d}{2}-1}(1-u)^{\frac{d}{2}-1} =
     \\
     =-\frac{\pi^{\frac{d}{2}}\Gamma(2-d/2)d}{|x-y|^{2-d}(d-2)}\int_0^1 du \:u^{\frac{d}{2}-1}(1-u)^{\frac{d}{2}-1} =
     \\
     = -\frac{\pi^{\frac{d}{2}}\Gamma(2-d/2)d}{|x-y|^{2-d}(d-2)} F(d),
\end{align}
where the following notation is introduced
\begin{equation}\label{F(d)}
     F(d) = \int_0^1 du\:u^{\frac{d}{2}-1}(1-u)^{\frac{d}{2}-1} = \mathrm{B} (d/2, d/2) = \frac{\Gamma(d/2)^2}{\Gamma(d)}.
\end{equation}
Using the identity
\begin{equation}
     \int_{0}^{1} du \: \log{u(1-u)} = -2,
\end{equation}
then for $d = 2 + \epsilon \to 2$
\begin{align}
     \int_0^1 du\:u^{\frac{d}{2}-1}(1-u)^{\frac{d}{2}-1} &= 1 + \frac{\epsilon}{ 2}\int_0^1 du\:\log(u(1-u)) + \mathcal{O}(\epsilon^2) = 1 - \epsilon + \mathcal{O}(\epsilon^2),
     \\
     \pi^{\frac{d}{2}} &= \pi\cdot \pi^{\frac{\epsilon}{2}} = \pi\left(1 + \frac{\epsilon}{2} \log(\pi) + \mathcal{O}(\epsilon^2)\right),
     \\
     \Gamma(2-d/2) &= \Gamma(1-\epsilon/2) = \Gamma(1)-\frac{\epsilon}{2}\Gamma'(1)+\mathcal{O}( \epsilon^2) = 1 + \frac{\epsilon}{2}\gamma_E+\mathcal{O}(\epsilon^2),
     \\
     |x-y|^{d-2} &= |x-y|^{\epsilon} = 1 +
     \epsilon \log|x-y| + \mathcal{O}(\epsilon^2),
\end{align}
where $\gamma_E$ is the Euler's constant. Thus, we have
\begin{multline}
     I_1^1(x,y) = -\frac{2\pi}{\epsilon}\biggl(1 + \frac{\epsilon}{2}\log\pi\biggr)\biggl(1 + \frac{ \epsilon}{2}\gamma_E\biggr)\biggl(1+\frac{\epsilon}{2}\biggr)\biggl(1 +
     \epsilon \log|x-y|\biggr)(1-\epsilon) + \mathcal{O}(\epsilon)=
     \\
     =-2\pi \biggl(\frac{1}{\epsilon} +\frac{1}{2}(\log{\pi} +\gamma_E +1 -2 + 2\log{|x-y|}) \biggr) = -\frac{2\pi}{\epsilon} - \pi\log{\bigl(|x-y|^2\pi e^{\gamma_E -1}\bigr)} + \mathcal{O} (\epsilon)=
     \\
     = -\frac{2\pi}{\epsilon} - \pi\log{\bigl(|x-y|^2 m_R^2\bigr)} + \mathcal{O}(\epsilon),
\end{multline}
where the following notation has been introduced
\begin{equation}
     m_R^2 = \pi e^{\gamma_E-1}.
\end{equation}

\section{Divergencies in factorised part of BB term}\label{int:fact}

Here we calculate all the divergencies of the factorised part of the BB term (\ref{eq:bb-term}).

\begin{equation}
\begin{split}
    \text{Fact} &= \int d^2w \:d^2z \:\sum_{i = 1}^{N}\left(\frac{\Delta_i}{(z - x_i)^2} + \frac{1}{z -x_i}\pdv{x_i} \right)\sum_{j = 1}^{N}\left(\frac{\Bar{\Delta}_j}{(\bar{z} - \bar{x}_j)^2} + \frac{1}{\bar{z} -\bar{x}_j}\pdv{\bar{x}_j} \right)\expval{T(w)\bar{T}(\bar{w})X}
    \\
    &= \int d^2w \:d^2z \:\sum_{i = 1}^{N}\frac{\Delta_i}{(z - x_i)^2}\sum_{j = 1}^{N}\frac{\Bar{\Delta}_j}{(\bar{z} - \bar{x}_j)^2}\sum_{l = 1}^{N}\left(\frac{\Delta_l}{(w - x_l)^2} + \frac{1}{w -x_l}\pdv{x_l} \right)\cdot
    \\
    &\cdot \sum_{k = 1}^{N}\left(\frac{\Bar{\Delta}_k}{(\bar{w} - \bar{x}_k)^2} + \frac{1}{\bar{w} -\bar{x}_k}\pdv{\bar{x}_k} \right)\expval{X}+
    \\
    &+ \int d^2w \:d^2z \:\sum_{i = 1}^{N}\frac{\Delta_i}{(z - x_i)^2}\sum_{j = 1}^{N}\frac{1}{\bar{z} -\bar{x}_j}\pdv{}{\bar{x}_j}\sum_{l = 1}^{N}\left(\frac{\Delta_l}{(w - x_l)^2} + \frac{1}{w -x_l}\pdv{x_l} \right)\cdot
    \\
  &\cdot \sum_{k = 1}^{N}\left(\frac{\Bar{\Delta}_k}{(\bar{w} - \bar{x}_k)^2} + \frac{1}{\bar{w} -\bar{x}_k}\pdv{\bar{x}_k} \right)\expval{X}
    + h.c. + \int d^2w \:d^2z \:\sum_{i = 1}^{N}\frac{1}{z - x_i}\pdv{}{x_i}\sum_{j = 1}^{N}\frac{1}{\bar{z} -\bar{x}_j}\pdv{}{\bar{x}_j}\cdot
    \\
  &\cdot \sum_{l = 1}^{N}\left(\frac{\Delta_l}{(w - x_l)^2} + \frac{1}{w -x_l}\pdv{x_l} \right) \sum_{k = 1}^{N}\left(\frac{\Bar{\Delta}_k}{(\bar{w} - \bar{x}_k)^2} + \frac{1}{\bar{w} -\bar{x}_k}\pdv{\bar{x}_k} \right)\expval{X}
\end{split}
\end{equation}
Let us introduce the notations
\begin{align}
    \text{Fact}_{22} &= \int d^2w \:d^2z \:\sum_{i = 1}^{N}\frac{\Delta_i}{(z - x_i)^2}\sum_{j = 1}^{N}\frac{\Bar{\Delta}_j}{(\bar{z} - \bar{x}_j)^2}\sum_{l = 1}^{N}\left(\frac{\Delta_l}{(w - x_l)^2} + \frac{1}{w -x_l}\pdv{x_l} \right)\cdot \nonumber
    \\
    &\cdot \sum_{k = 1}^{N}\left(\frac{\Bar{\Delta}_k}{(\bar{w} - \bar{x}_k)^2} + \frac{1}{\bar{w} -\bar{x}_k}\pdv{\bar{x}_k} \right)\expval{X}
\end{align}
\begin{align}
    \text{Fact}_{21} &= \int d^2w \:d^2z \:\sum_{i = 1}^{N}\frac{1}{(z - x_i)^2}\sum_{j = 1}^{N}\frac{1}{\bar{z} -\bar{x}_j}\pdv{}{\bar{x}_j}\sum_{l = 1}^{N}\left(\frac{\Delta_l}{(w - x_l)^2} + \frac{1}{w -x_l}\pdv{x_l} \right)\cdot\nonumber
    \\
    &\cdot \sum_{k = 1}^{N}\left(\frac{\Bar{\Delta}_k}{(\bar{w} - \bar{x}_k)^2} + \frac{1}{\bar{w} -\bar{x}_k}\pdv{\bar{x}_k} \right)\expval{X}
\end{align}
\begin{align}
    \text{Fact}_{12} &= \int d^2w \:d^2z \:\sum_{i = 1}^{N}\frac{1}{(\bar{z} - \bar{x}_i)^2}\sum_{j = 1}^{N}\frac{1}{z -x_j}\pdv{}{x_j}\sum_{l = 1}^{N}\left(\frac{\Delta_l}{(w - x_l)^2} + \frac{1}{w -x_l}\pdv{x_l} \right)\cdot\nonumber
    \\
    &\cdot \sum_{k = 1}^{N}\left(\frac{\Bar{\Delta}_k}{(\bar{w} - \bar{x}_k)^2} + \frac{1}{\bar{w} -\bar{x}_k}\pdv{\bar{x}_k} \right)\expval{X}
\end{align}
\begin{align}
    \text{Fact}_{11} &= \int d^2w \:d^2z \:\sum_{i = 1}^{N}\frac{1}{z - x_i}\pdv{}{x_i}\sum_{j = 1}^{N}\frac{1}{\bar{z} -\bar{x}_j}\pdv{}{\bar{x}_j}\cdot\nonumber
    \\
    &\cdot \sum_{l = 1}^{N}\left(\frac{\Delta_l}{(w - x_l)^2} + \frac{1}{w -x_l}\pdv{x_l} \right) \sum_{k = 1}^{N}\left(\frac{\Bar{\Delta}_k}{(\bar{w} - \bar{x}_k)^2} + \frac{1}{\bar{w} -\bar{x}_k}\pdv{\bar{x}_k} \right)\expval{X}
\end{align}
Then we have
\begin{equation}
    \text{Fact} = \text{Fact}_{22}+\text{Fact}_{21}+\text{Fact}_{12}+\text{Fact}_{11}.
\end{equation}
As $I_2^2 = 0 + \mathcal{O}(\epsilon)$, then only the poles of the integration over $w$ contribute to $\text{Fact}_{22}$.
\begin{equation}
\begin{split}
    \text{Fact}_{22} &= \int d^2w \:d^2z \:\sum_{i = 1}^{N}\frac{\Delta_i}{(z - x_i)^2}\sum_{j = 1}^{N}\frac{\Bar{\Delta}_j}{(\bar{z} - \bar{x}_j)^2}\sum_{l = 1}^{N}\left(\frac{\Delta_l}{(w - x_l)^2} + \frac{1}{w -x_l}\pdv{x_l} \right)
    \\
    &\cdot \sum_{k = 1}^{N}\left(\frac{\Bar{\Delta}_k}{(\bar{w} - \bar{x}_k)^2} + \frac{1}{\bar{w} -\bar{x}_k}\pdv{\bar{x}_k} \right)\expval{X}
    \\
    &= \sum_{i = 1, \: j \neq i}^N \Delta_i\bar{\Delta}_j \left(0 + \frac{\pi}{2|x_i-x_j|^2} \epsilon\right)\sum_{l = 1, \: k \neq l}^N \left(-\frac{2\pi}{\epsilon}\right)\pdv{x_l}\pdv{\bar{x}_k}\expval{X} =
    \\
    &= -\pi^2 \sum_{i = 1, \: j \neq i}^N \sum_{l = 1, \: k \neq l}^N \frac{\Delta_i\bar{\Delta}_j}{|x_i-x_j|^2}\pdv{x_l}\pdv{\bar{x}_k}\expval{X}.
\end{split}
\end{equation}
Thus we conclude that $\text{Fact}_{22}$ is finite.

Now in order to calculate the divergences in $\text{Fact}_{21}$ we use the expansion
\begin{equation}
    I_2^1(x, y) = -\frac{\pi}{x-y} - \frac{\pi\log m|x-y| }{x-y}\epsilon.
\end{equation}
It means that integrals over $w$ having an expansion of the form $0 + \mathcal{\epsilon}$ will not contribute.
\begin{equation}
\begin{split}
    \text{Fact}_{21} &= \int d^2w \:d^2z \:\sum_{i = 1}^{N}\frac{1}{(z - x_i)^2}\sum_{j = 1}^{N}\frac{1}{\bar{z} -\bar{x}_j}\pdv{}{\bar{x}_j}\sum_{l = 1}^{N}\left(\frac{\Delta_l}{(w - x_l)^2} + \frac{1}{w -x_l}\pdv{x_l} \right)\cdot
    \\
    &\cdot \sum_{k = 1}^{N}\left(\frac{\Bar{\Delta}_k}{(\bar{w} - \bar{x}_k)^2} + \frac{1}{\bar{w} -\bar{x}_k}\pdv{\bar{x}_k} \right)\expval{X} 
    \\
    &=-\pi^2\sum_{j,\: i \neq j}\Delta_i\left(\frac{1}{x_i-x_j}+\frac{\log{m|x_i-x_j|}}{x_i-x_j}\epsilon\right)\bar{\partial}_j
    \\
    &\sum_{l, \: k \neq l} \biggl[-\frac{\Delta_l}{x_l-x_k}\bar{\partial}_k + \frac{\bar{\Delta}_k}{\bar{x}_l-\bar{x}_k}\partial_l - \left(\frac{2}{\epsilon} + \log m^2|x_l-x_k|^2\right)\partial_l\bar{\partial}_k\biggr]\expval{X}
\end{split}
\end{equation}
Note that
\begin{equation}
    \sum_{l, \: k \neq l} = \sum_{l, \: k\neq j,\: k \neq l} + \sum_{k\neq j} (l = j) + \sum_{l \neq j} (k = j)
\end{equation}

In the integral $\text{Fact}_{11}$ we select the divergent terms. First, there will be a second-order pole associated with the multiplication of first-order poles:
\begin{multline}
    \int d^2w \: d^2z \: \sum_{i, j}^{N} \frac{1}{(z-x_i)(\bar{z} - \bar{x}_j)}\sum_{l \neq i, k \neq j}^N \frac{1}{(w-x_l)(\bar{w}-\bar{x}_k)} \partial_i \bar{\partial}_j \partial_l \bar{\partial}_k\expval{X} \rightarrow
    \\
    \rightarrow \left(\frac{2\pi}{\epsilon}\right)^2\sum_{i, j \neq i}^{N} \: \sum_{l \neq i, k \neq j, l \neq k}^N \partial_i \bar{\partial}_j \partial_l \bar{\partial}_k\expval{X}
\end{multline}
First-order poles in $\text{Fact}_{11}$:
\begin{multline}
    \int d^2w \: d^2z \:\sum_{i, j}^{N}\frac{1}{(z - x_i)(\bar{z} -\bar{x}_j)}\partial_i\partial_j \sum_{l, k}^{N}\left(\frac{\Delta_l}{(w - x_l)^2} + \frac{1}{w -x_l}\partial_l \right) \left(\frac{\Bar{\Delta}_k}{(\bar{w} - \bar{x}_k)^2} + \frac{1}{\bar{w} -\bar{x}_k}\bar{\partial}_k \right)\expval{X} \rightarrow
    \\
    \rightarrow \sum_{i, j \neq i}^{N} \left(-\frac{2\pi}{\epsilon} - 2 \pi \log(m|x_i - x_j|)\right)\partial_i\bar{\partial}_j \: \cdot \\
    \cdot \pi \sum_{l, k \neq l}^{N}\left(\frac{\bar{\Delta}_k}{\bar{x}_l-\bar{x}_k}\partial_l - \frac{\Delta_l}{x_l - x_k}\bar{\partial}_k + \left(-\frac{2\pi}{\epsilon} - 2 \pi \log(m|x_l - x_k|)\right)\partial_l\bar{\partial}_k\right)\expval{X} \rightarrow
    \\
    \rightarrow -\frac{2\pi^2}{\epsilon}\sum_{i, j \neq i} \sum_{l, k \neq l} \left(\frac{\bar{\Delta}_k}{\bar{x}_l-\bar{x}_k}\partial_l - \frac{\Delta_l}{x_l - x_k}\bar{\partial}_k - 2 \log(m|x_l - x_k|)\partial_l\bar{\partial}_k\right)\partial_i\bar{\partial}_j\expval{X} + 
    \\
    + \frac{(2 \pi)^2}{\epsilon}\sum_{i, j \neq i} \sum_{l, k \neq l}\log(m|x_i - x_j|)\partial_i \bar{\partial}_j \partial_l \bar{\partial}_k\expval{X} =
    \\
    =-\frac{2\pi^2}{\epsilon}\sum_{i, j \neq i} \sum_{l, k \neq l} \left(\frac{\bar{\Delta}_k}{\bar{x}_l-\bar{x}_k}\partial_l - \frac{\Delta_l}{x_l - x_k}\bar{\partial}_k - 4 \log(m|x_l - x_k|)\partial_l\bar{\partial}_k\right)\partial_i\bar{\partial}_j\expval{X}
\end{multline}
These are all divergences from the 'factorized'\:part:
\begin{equation}
    \begin{split}
        \frac{2 \pi^2}{\epsilon}\sum_{i,k, j \neq i, l \neq k} \frac{\Delta_i}{x_i-x_j}\bar{\partial}_j\partial_l\bar{\partial}_k\expval{X} -\frac{2 \pi^2}{\epsilon}\sum_{i,k, j \neq i, l \neq k} \frac{\bar{\Delta}_i}{\bar{x}_i-\bar{x}_j}\partial_j\partial_l\bar{\partial}_k\expval{X}
        \\
        -\frac{2\pi^2}{\epsilon}\sum_{i, j \neq i} \sum_{l, k \neq l} \left(\frac{\bar{\Delta}_k}{\bar{x}_l-\bar{x}_k}\partial_l - \frac{\Delta_l}{x_l - x_k}\bar{\partial}_k - 4 \log(m|x_l - x_k|)\partial_l\bar{\partial}_k\right)\partial_i\bar{\partial}_j\expval{X}
        \\
         +\left(\frac{2\pi}{\epsilon}\right)^2\sum_{i; j \neq i;l \neq i,k; k \neq j}^N \partial_i \bar{\partial}_j \partial_l \bar{\partial}_k\expval{X}
    \end{split}
\end{equation}

\bibliographystyle{JHEP}
\bibliography{biblio.bib}





\end{document}